\documentclass[
  aps,
  prd,
  twocolumn,
  superscriptaddress,
  nofootinbib,
  longbibliography,
  showkeys
]{revtex4-2}

\usepackage{amsmath,amssymb,mathtools,bm,amsthm}
\usepackage{graphicx}
\usepackage{booktabs}
\usepackage{microtype}
\usepackage{xcolor}
\usepackage{hyperref}
\usepackage{url}

\hypersetup{
  colorlinks=true,
  linkcolor=blue!45!black,
  citecolor=blue!45!black,
  urlcolor=blue!45!black,
  pdftitle={Relational Wigner--Smith Duration in Wheeler--DeWitt Scattering},
  pdfauthor={Daulet Berkimbayev},
  pdfsubject={Relational duration and local clocks in Wheeler--DeWitt quantum cosmology},
  pdfkeywords={Wheeler--DeWitt equation, relational time, Wigner--Smith operator,
    quantum cosmology, internal clocks, multichannel scattering}
}

\theoremstyle{remark}

\newcommand{\dd}{\mathrm d}
\newcommand{\I}{\mathbb I}
\newcommand{\Hphys}{\mathcal H_{\mathrm{phys}}}
\newcommand{\Tr}{\operatorname{Tr}}

\newcommand{\sech}{\operatorname{sech}}

\newcommand{\erfc}{\operatorname{erfc}}
\newcommand{\Qint}{Q_{\mathrm{int}}}
\newcommand{\Qsec}{Q_{\mathrm{sec}}}
\newcommand{\Pexc}{P_{\mathrm{exc}}}
\newcommand{\Sent}{S_{\mathrm{sp}}}

\newcommand{\missingfigure}[1]{%
  \fbox{\parbox[c][3.0cm][c]{0.94\linewidth}{%
  \centering\small Figure not found. Place the established figures in
  \texttt{paper2\_figures}; generate the new figure with
  \texttt{article2\_new\_results\_figure.py}.\\[2mm]#1}}}

\begin{document}

\title{Relational Wigner--Smith Duration in Wheeler--DeWitt Scattering}

\author{Daulet Berkimbayev}
\email{daulet9432@gmail.com}
\affiliation{Al-Farabi Kazakh National University, Al-Farabi av. 71, Almaty 050040, Kazakhstan}

\begin{abstract}
A closed Friedmann--Lema\^{i}tre--Robertson--Walker Wheeler--DeWitt model is
formulated as an exact reflection problem.  The derivative of its reflection
phase defines a relational crossing duration whose first two clock moments
follow from a covariant scalar-clock observable.  An analytic expression is
obtained for this duration, its classical recollapse limit, and the leading
quantum correction.  A finite spectral packet also gives an operational, equal-prior minimum error probability for distinguishing the two orientations of
recollapse with a geometric reading.  The construction is extended to a
bounded finite quantum detector.  Its multichannel reflection matrix yields
probe transitions, spectral--probe correlations, and a matrix duration. A
direct weak-coupling calculation verifies the predicted transition and
duration scalings.  Scalar-clock conditioning and oriented geometric sections
are shown to be local representations of the same positive-frequency Dirac
sector.  These results define duration and transition observables entirely
through correlations and scattering records of a stationary constrained
state, without introducing a background clock.
\end{abstract}


\maketitle

\section{Introduction}
\label{sec:introduction}

Canonical metric quantum gravity replaces ordinary Hamiltonian evolution by
constraints on physical states, so  in the Wheeler--DeWitt formulation,
\begin{equation}
 \widehat{\mathcal C}\Psi=0,
 \label{eq:abstract_wdw}
\end{equation}
therefore predictions must be expressed as correlations among physical degrees of
freedom rather than as evolution relative to an externally prescribed
parameter \cite{DeWitt1967,Isham1993,Anderson2012}.  Page--Wootters
conditioning \cite{PageWootters1983}, evolving constants and partial
observables \cite{Rovelli1991,Rovelli2002}, complete observables
\cite{Dittrich2007}, and clock-neutral reductions
\cite{Hohn2019,HohnVanrietvelde2020,HohnSmithLock2021} provide complementary
realizations of this relational idea.

Exact minisuperspace wave packets were constructed early in quantum cosmology
\cite{Kiefer1988}.  Matter transitions and backscattering between
expanding and contracting Wheeler--DeWitt components were studied by Massar
and Parentani \cite{MassarParentani1998,MassarParentani1999}, and subsequent
work analysed transitions near cosmological turning regions
\cite{Coutant2014,Coutant2016}.  Relational evolution in a recollapsing
universe is state dependent and may fail before a classical turning point
because different spectral components turn at different positions
\cite{HohnKubalovaTsobanjan2012}.  Clock dependence in quantum recollapse has
also been demonstrated explicitly \cite{Gielen2020,Gielen2022}.  Recent
studies formulate Wheeler--DeWitt cosmology directly as relativistic
scattering between cosmological branches
\cite{GiovannettiMaioneMontani2023,LoFrancoMontani2024,
Giovannetti2025,LoFrancoMontani2026}.  Covariant clock POVMs and the
equivalence of Dirac, Page--Wootters, and reduced descriptions are therefore
used here as established tools rather than claimed as new general
principles \cite{HohnSmithLock2021}.

Phase time has also been reconsidered as a route to positive probabilities
and relational Ehrenfest relations in constrained systems
\cite{Chataignier2024}.  The observable constructed below is different: it
is the Wigner--Smith derivative of a reflection operator
\cite{Wigner1955,Smith1960}, fixed operationally by the difference between
incoming and outgoing readings of the same covariant scalar clock.

The advance of the present calculation is more specific.  First, the exact
closed-FLRW reflection phase is converted into a Wigner--Smith crossing
duration and related to the first two moments of a covariant clock
distribution.  The result is compared analytically with the classical
recollapse duration, exposing the leading quantum correction.  The
construction is tied directly to an internal clock and then extended to a
matrix reflection operator.  Second,
the loss of an unoriented geometric clock is expressed as a minimum
branch-discrimination error rather than by a chosen peak-separation
threshold.  Third, an exact finite detector supplies a controlled
multichannel realization in which the transition and duration scalings are
derived directly.

The calculation concerns a homogeneous
minisuperspace and a bounded detector of fixed dimension.  It neither claims
a limit to an unrestricted matter field nor addresses the local constraint
algebra of field theory.  Within this sector, however, crossing duration,
channel transfer, and reduced detector records are defined without adding an
independent background time.

\section{Relational duration}
\label{sec:one_channel}

In units $8\pi G=\hbar=1$, let $\alpha=\ln a$ be the logarithmic scale factor
of a closed FLRW geometry and let $\tau$ be a rescaled homogeneous massless
scalar.  A convenient densitized constraint is
\begin{equation}
 {\mathcal C}_0=-p_\alpha^2+p_\tau^2-36e^{4\alpha}\approx0 .
 \label{eq:background_constraint}
\end{equation}
Laplace--Beltrami ordering gives
\begin{equation}
 \left[\partial_\alpha^2-\partial_\tau^2-36e^{4\alpha}\right]
 \Psi(\alpha,\tau)=0 .
 \label{eq:background_wdw}
\end{equation}
On the positive scalar-frequency sector,
\begin{equation}
 -i\partial_\tau\Psi=\sqrt{\Theta_0}\Psi,
 \qquad
 \Theta_0=-\partial_\alpha^2+36e^{4\alpha},
 \label{eq:positive_frequency}
\end{equation}
where spectral components are proportional to $e^{+i\omega\tau}$.

The normalized generalized eigenfunctions are
\begin{equation}
 e_\omega(\alpha)=
 \frac{\sqrt{\omega\sinh(\pi\omega/2)}}{\pi}
 K_{i\omega/2}\!\left(3e^{2\alpha}\right),
 \qquad \omega>0,
 \label{eq:bessel_modes}
\end{equation}
and a physical packet is
\begin{align}
 \Psi_A(\alpha,\tau)
 &=\int_0^\infty\dd\omega\,
 A(\omega)e_\omega(\alpha)e^{i\omega\tau},\\
 \|A\|^2&=\int_0^\infty\dd\omega\,|A(\omega)|^2=1 .
 \label{eq:background_packet}
\end{align}
Group averaging on this sector gives
\begin{equation}
 \Hphys\simeq L^2(\mathbb R_+,\dd\omega).
 \label{eq:physical_space}
\end{equation}
Because the frequency is semibounded, scalar readings are represented by the
covariant POVM
\begin{align}
 (\tau|\omega\rangle&=\frac{e^{i\omega\tau}}{\sqrt{2\pi}},\\
 E_\tau(\dd\tau)&=|\tau)(\tau|\dd\tau,\\
 \int_{-\infty}^{\infty}E_\tau(\dd\tau)&=\I_{\Hphys}.
 \label{eq:scalar_clock_povm}
\end{align}
This supplies the clock statistics without postulating a self-adjoint
operator canonically conjugate to a semibounded Hamiltonian
\cite{DasNoth2021,HohnSmithLock2021}.

The corresponding classical relational solution is
\begin{align}
 e^{2\alpha(\tau)}
 &=\frac{\omega}{6}\sech\!\left[2(\tau-\tau_0)\right],\\
 \alpha_{\rm t}(\omega)&=\frac12\ln\frac{\omega}{6} .
 \label{eq:classical_trajectory}
\end{align}
Every $\alpha<\alpha_{\rm t}$ is crossed once on each orientation.

At the open endpoint $\alpha\rightarrow-\infty$, Eq.~\eqref{eq:bessel_modes}
has the exact asymptotic form
\begin{equation}
 e_\omega(\alpha)\sim\frac{1}{\sqrt{2\pi}}
 \left[e^{i\delta(\omega)}e^{-i\omega\alpha}
 +e^{-i\delta(\omega)}e^{+i\omega\alpha}\right],
 \label{eq:background_asymptotic}
\end{equation}
where
\begin{equation}
 \delta(\omega)=\arg\Gamma\!\left(\frac{i\omega}{2}\right)
 -\frac{\omega}{2}\ln\frac32 .
 \label{eq:phase_shift}
\end{equation}
The reflection operator is therefore
\begin{equation}
 S_0(\omega)=e^{-2i\delta(\omega)}
 =\frac{\Gamma(-i\omega/2)}{\Gamma(i\omega/2)}
 \left(\frac32\right)^{i\omega},
 \qquad |S_0|=1 .
 \label{eq:exact_s}
\end{equation}

It is important to propose that the continuum of $\Theta_0$ selected by decay at
$\alpha\rightarrow+\infty$ has spectral multiplicity one.  Consequently,
no two nonzero orthogonal projectors can commute with $\sqrt{\Theta_0}$,
sum to the identity on a continuum interval, and distinguish expanding and
contracting directions at the same frequency.

The proposition is narrow as it only states that these orientations are not two independent conserved stationary sectors of the selected one-channel theory.  Appendix~\ref{app:proofs} gives the derivation.

At a reference section $\alpha_0$ in the open region, free propagation phases
give
\begin{equation}
 S_{\alpha_0}(\omega)
 =e^{-2i\delta(\omega)+2i\omega\alpha_0} .
 \label{eq:reference_s}
\end{equation}
For the $e^{+i\omega\tau}$ convention, define
\begin{equation}
 Q_0(\alpha_0)
 =iS_{\alpha_0}^\dagger\partial_\omega S_{\alpha_0}
 =2\left[\delta'(\omega)-\alpha_0\right] .
 \label{eq:wigner_smith_one}
\end{equation}
The sign differs from the usual $e^{-iE t}$ scattering convention and is
fixed here by the scalar-clock Fourier transform.

Let $A$ be normalized, absolutely continuous, and vanish at the endpoints of
its spectral support.  The incoming and outgoing scalar-clock distributions
at $\alpha_0$ satisfy
\begin{equation}
 \langle\tau\rangle_{\rm out}-\langle\tau\rangle_{\rm in}
 =\langle A|Q_0(\alpha_0)|A\rangle .
 \label{eq:duration_mean}
\end{equation}
Writing $T=i\partial_\omega$ on this domain, their variances obey
\begin{equation}
 \operatorname{Var}_{\rm out}(\tau)
 =\operatorname{Var}_{A}(T+Q_0),
 \label{eq:duration_variance_general}
\end{equation}
or equivalently
\begin{align}
 \operatorname{Var}_{\rm out}(\tau)-\operatorname{Var}_{\rm in}(\tau)
 =&\operatorname{Var}_{A}(Q_0)
 +\langle\{\Delta T,\Delta Q_0\}\rangle_A .
 \label{eq:duration_variance}
\end{align}
For a real spectral envelope, the covariance term vanishes and the added
clock variance is exactly $\operatorname{Var}_{A}(Q_0)$.

The framework follows from the operator identity
$S_{\alpha_0}^\dagger T S_{\alpha_0}=T+Q_0$; no semiclassical approximation
is used.  For a real narrow Gaussian of frequency variance
$\sigma_\omega^2$,
\begin{align}
 \operatorname{Var}_{\rm in}(\tau)&=\frac{1}{4\sigma_\omega^2},\\
 \operatorname{Var}_{\rm out}(\tau)
 &=\frac{1}{4\sigma_\omega^2}
 +Q_0'(\omega_0)^2\sigma_\omega^2
 +O(\sigma_\omega^4).
 \label{eq:narrow_variance}
\end{align}

The exact phase makes the mean duration analytic:
\begin{equation}
 Q_0(\omega;\alpha_0)
 =\operatorname{Re}\psi\!\left(\frac{i\omega}{2}\right)
 -\ln\frac32-2\alpha_0 ,
 \label{eq:exact_duration}
\end{equation}
where $\psi$ is the digamma function.  The classical scalar-clock time between
the two crossings of the same section follows from
Eq.~\eqref{eq:classical_trajectory}:
\begin{equation}
 \Delta\tau_{\rm cl}(\omega;\alpha_0)
 =\operatorname{arcosh}\!\left(
 \frac{\omega}{6e^{2\alpha_0}}
 \right).
 \label{eq:classical_duration}
\end{equation}

At fixed $\alpha_0$ and large $\omega$,
\begin{align}
 Q_0
 &=\ln\!\frac{\omega}{3e^{2\alpha_0}}
 +\frac{1}{3\omega^2}+O(\omega^{-4}),\\
 \Delta\tau_{\rm cl}
 &=\ln\!\frac{\omega}{3e^{2\alpha_0}}
 -\frac{9e^{4\alpha_0}}{\omega^2}+O(\omega^{-4}),
 \label{eq:duration_expansions}
\end{align}
and hence
\begin{equation}
 Q_0-\Delta\tau_{\rm cl}
 =\frac{\tfrac13+9e^{4\alpha_0}}{\omega^2}
 +O(\omega^{-4}).
 \label{eq:duration_correction}
\end{equation}

Thus the Wigner--Smith observable, while being a formal phase derivative, also recovers the classical relational duration and supplies its leading quantum
correction.  The left panel of Fig.~\ref{fig:new_diagnostics} displays the
exact comparison and the scaled residual.

The packet also determines where an unoriented geometric reading becomes
ambiguous.  A purely kinematic indicator is the classically allowed packet
fraction
\begin{equation}
 F_{\rm allow}(\alpha)
 =\int_0^\infty\dd\omega\,|A(\omega)|^2
 \mathbf 1_{\{\omega>6e^{2\alpha}\}} .
 \label{eq:allowed_fraction}
\end{equation}
Since $\alpha_{\rm t}(\omega)$ is monotonic,
\begin{equation}
 F_{\rm allow}(\alpha)=\Pr(\alpha_{\rm t}>\alpha).
 \label{eq:turning_survival}
\end{equation}
This identity locates the distributed turning region but does not by itself
measure clock quality.

An operational measure follows by asking how well the scalar-clock record
distinguishes the two orientations at fixed $\alpha$.  Put
\begin{align}
 h(\alpha)&=6e^{2\alpha},&
 u(\omega,\alpha)&=\operatorname{arcosh}\frac{\omega}{h(\alpha)} .
\end{align}
Away from the Airy region, the two local WKB phases are
\begin{equation}
 \vartheta_\pm(\omega,\alpha)
 =\pm\left\{\frac{\omega}{2}
 [u-\tanh u]+\frac{\pi}{4}\right\} .
 \label{eq:wkb_branch_phase}
\end{equation}
Their clock distributions have centers separated by
$u(\omega_0,\alpha)=\Delta\tau_{\rm cl}$ because
$\partial_\omega\vartheta_\pm=\pm u/2$.

For a narrow Gaussian spectral density of variance $\sigma_\omega^2$, each
branch distribution is Gaussian to leading order, with
\begin{equation}
 \sigma_\tau^2(\alpha)
 =\frac{1}{4\sigma_\omega^2}
 +\frac{\sigma_\omega^2}
 {4[\omega_0^2-h(\alpha)^2]} .
 \label{eq:branch_clock_width}
\end{equation}
The first term is the Fourier-limited clock width and the second is the WKB
phase-curvature contribution.  For equal priors, the minimum error attainable
from the scalar reading is the Bayes error
\begin{equation}
 P_{\rm err}(\alpha)
 =\frac12\erfc\!\left[
 \frac{u(\omega_0,\alpha)}{2\sqrt{2}\,\sigma_\tau(\alpha)}
 \right].
 \label{eq:branch_error}
\end{equation}
Unlike a peak-separation threshold, Eq.~\eqref{eq:branch_error} is an
operational probability.  It approaches zero for well separated clock
records and $1/2$ when no orientation information can be extracted.  The
linearized turning equation has Airy length
\begin{equation}
 \ell_{\rm A}=(4\omega_0^2)^{-1/3},
 \label{eq:airy_length}
\end{equation}
which marks the domain where the WKB branch split itself ceases to be a local
description.  The center panel of Fig.~\ref{fig:new_diagnostics} compares
Eq.~\eqref{eq:branch_error} with the total-variation error obtained from the
two WKB clock distributions before that region.

\begin{figure*}[t]
 \centering
 \IfFileExists{paper2_figures/fig1_one_channel_structure.pdf}
 {\includegraphics[width=\textwidth]{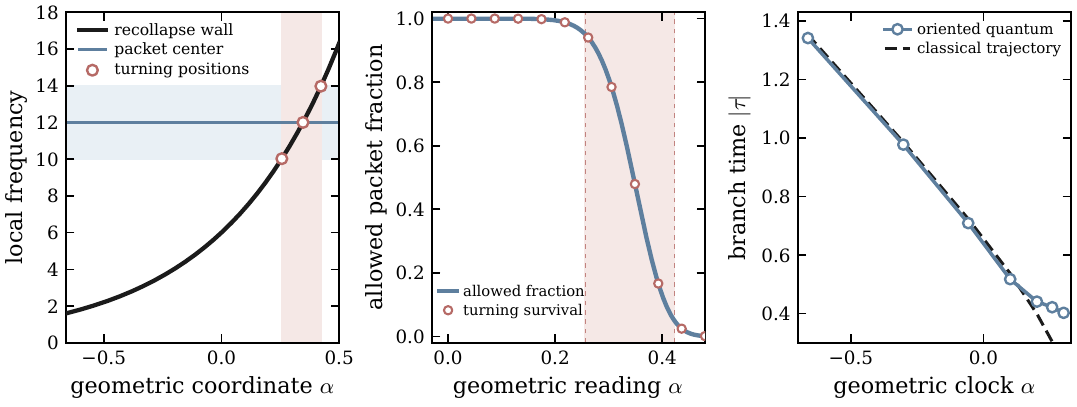}}
 {\missingfigure{One-channel wall, turning distribution, and oriented range.}}
 \caption{One-channel packet structure.  Left: the exact recollapse wall and
 the packet's distributed turning positions.  Center: the allowed packet
 fraction equals the survival function of the turning-position distribution,
 Eq.~\eqref{eq:turning_survival}.  Right: an orientation record distinguishes
 local branches beyond the range of an unoriented reading.}
 \label{fig:one_channel_structure}
\end{figure*}

\begin{figure*}[t]
 \centering
 \IfFileExists{paper2_figures/fig2_exact_packet.pdf}
 {\includegraphics[width=\textwidth]{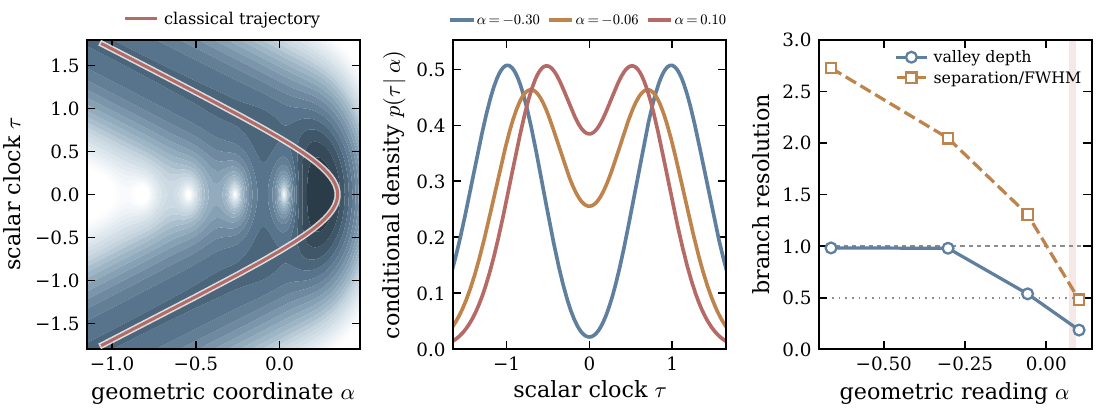}}
 {\missingfigure{Exact Bessel packet and conditional scalar-clock records.}}
 \caption{Exact finite-width Bessel packet.  Left: configuration-space
 density with the classical relational trajectory.  Center: conditional
 scalar-clock distributions at three geometric readings; their two peaks
 merge as recollapse is approached.  Right: valley-depth and
 separation-to-FWHM curves provide a visual comparison with the operational
 error probability.}
 \label{fig:exact_packet}
\end{figure*}

Figures~\ref{fig:one_channel_structure} and \ref{fig:exact_packet} retain the
complete packet picture.  They are now used as visualizations of the turning
distribution and the exact conditional density.

\section{Finite-detector multichannel reflection}
\label{sec:probe}

The one-channel model is coupled next to a bounded detector of fixed
dimension $M$.  In the even oscillator-inspired basis
$|r\rangle$, $r=0,2,\ldots,2(M-1)$, define
\begin{align}
 \bm H_{\rm p}(\alpha)
 &=n\bm D_M+\frac{\mu^2}{2}e^{2\alpha}\bm V_M,\\
 (D_M)_{rs}&=\left(r+\frac12\right)\delta_{rs},
 \label{eq:finite_probe_hamiltonian}
\end{align}
with
\begin{align}
 (V_M)_{rs}=\frac{1}{2n}\Bigl[&
 \sqrt{s(s-1)}\,\delta_{r,s-2}+(2s+1)\delta_{rs}\nonumber\\[-1mm]
 &+\sqrt{(s+1)(s+2)}\,\delta_{r,s+2}\Bigr].
 \label{eq:v2}
\end{align}
These finite matrices specify the detector.

Adding its energy before the same positive densitization as in
Eq.~\eqref{eq:background_constraint} gives
\begin{equation}
 \mathcal C_M=-p_\alpha^2+p_\tau^2-36e^{4\alpha}
 +12e^{2\alpha}\bm H_{\rm p}(\alpha)\approx0 .
 \label{eq:coupled_constraint}
\end{equation}
The coupled WDW equation is
\begin{equation}
 \left[\partial_\alpha^2-\partial_\tau^2-36e^{4\alpha}\I_M
 +12e^{2\alpha}\bm H_{\rm p}(\alpha)\right]\bm\Psi=0 .
 \label{eq:coupled_wdw}
\end{equation}
On its positive scalar-frequency sector,
\begin{align}
 \Hphys^{(M)}&\simeq
 L^2(\mathbb R_+,\dd\omega)\otimes\mathbb C^M,\\
 \Theta_M&=-\I_M\partial_\alpha^2+\bm W_M(\alpha),\\
 \bm W_M(\alpha)
 &=6e^{4\alpha}\bm B_M-12ne^{2\alpha}\bm D_M,\\
 \bm B_M&=6\I_M-\mu^2\bm V_M .
 \label{eq:matrix_operator}
\end{align}

The detector model confines every internal direction if
\begin{equation}
 b_M:=\lambda_{\min}(\bm B_M)
 =6-\mu^2\lambda_{\max}(\bm V_M)>0 .
 \label{eq:wall_theorem}
\end{equation}
For $b_M>0$, $\Theta_M$ is semibounded and has a unique self-adjoint
limit-point realization.  Its essential spectrum is
$[0,\infty)$ with continuum multiplicity $M$.
\label{thm:wall}

Appendix~\ref{app:wall} provides steps of this statement using the exact largest
eigenvalue of $\bm V_M$. Negative discrete eigenvalues, if present, lie outside
the selected oscillatory positive-frequency sector.

At the open endpoint,
\begin{equation}
 \bm\chi_\omega(\alpha)\sim
 \frac{e^{-i\omega\alpha}}{\sqrt{2\omega}}\bm a_{\rm in}
 +\frac{e^{+i\omega\alpha}}{\sqrt{2\omega}}\bm a_{\rm out},
\end{equation}
and the confining boundary condition defines a unitary reflection matrix
\begin{equation}
 \bm a_{\rm out}=\bm S_M(\omega)\bm a_{\rm in},
 \quad \bm S_M^\dagger\bm S_M=\I_M .
 \label{eq:matrix_s}
\end{equation}
For ground-channel input $\bm e_0$ and packet $A$, let
\begin{equation}
 \bm c_{\rm out}(\omega)=A(\omega)\bm S_M(\omega)\bm e_0 .
 \label{eq:outgoing_spectral_amplitude}
\end{equation}
The asymptotic detector probabilities, reduced state, and
spectral--detector entropy are
\begin{align}
 P_r&=\int_0^\infty\dd\omega\,|c_{{\rm out},r}(\omega)|^2,\\
 \bm\rho_{\rm p}^{\rm out}
 &=\int_0^\infty\dd\omega\,
 \bm c_{\rm out}(\omega)\bm c_{\rm out}^\dagger(\omega),\\
 \Sent&=-\Tr\left(\bm\rho_{\rm p}^{\rm out}
 \ln\bm\rho_{\rm p}^{\rm out}\right).
 \label{eq:probe_observables}
\end{align}
Here $\Sent$ quantifies correlations within the finite asymptotic Hilbert
space.

The weak-coupling hierarchy follows before numerical integration.  Write
$\bm W_M=\bm W_0+\mu^2\bm W_1$, with
$\bm W_1(\alpha)=-6e^{4\alpha}\bm V_M$.  Away from thresholds, ordinary
parameter-dependent scattering theory gives
\begin{equation}
 S_{r0}(\omega;\mu)=\delta_{r0}S_r^{(0)}(\omega)
 +\mu^2 {\cal A}_{r0}(\omega)+O(\mu^4),
 \label{eq:born_expansion}
\end{equation}
where ${\cal A}_{r0}$ is the corresponding distorted-wave matrix element of
$\bm W_1$.  The selection rule in Eq.~\eqref{eq:v2} makes $r=2$ the leading
excited channel.

For ground-channel incidence away from a threshold,
\begin{align}
 \Pexc(\omega)&=C_P(\omega)\mu^4+O(\mu^6),\\
 Q_{00}(\omega;\mu)-Q_{00}(\omega;0)
 &=C_Q(\omega)\mu^2+O(\mu^4).
 \label{eq:weak_scaling}
\end{align}
The first coefficient is the sum of the squared off-diagonal amplitudes in
Eq.~\eqref{eq:born_expansion}; the second is the frequency
derivative of the diagonal first-order phase.

\begin{figure*}[t]
 \centering
 \IfFileExists{paper2_figures/fig_new_relational_diagnostics.pdf}
 {\includegraphics[width=\textwidth]{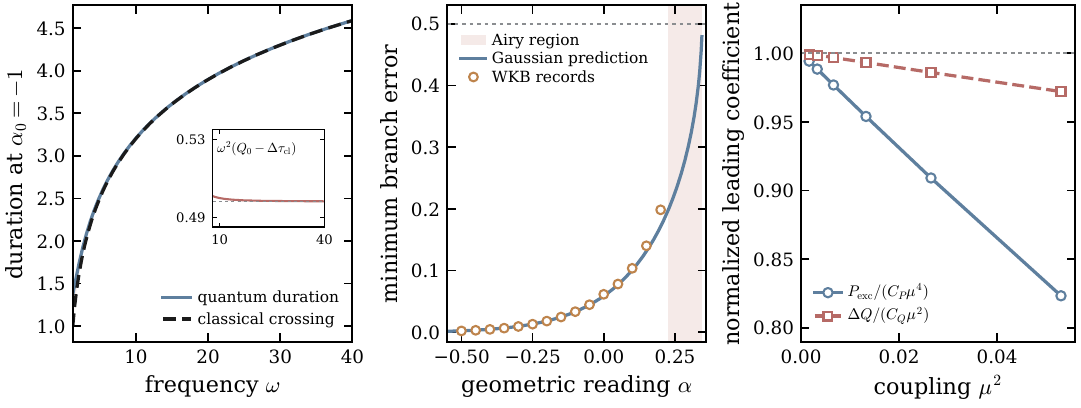}}
 {\missingfigure{New duration, branch-error, and weak-coupling diagnostics.}}
 \caption{Left: the exact
 Wigner--Smith duration \eqref{eq:exact_duration} approaches the classical
 two-crossing duration \eqref{eq:classical_duration}; the inset shows the
 scaled $O(\omega^{-2})$ correction.  Center: the minimum error for
 distinguishing the two scalar-clock records follows
 Eq.~\eqref{eq:branch_error}; open markers are obtained by Fourier transforming
 the two WKB branch amplitudes, and the shaded interval is the Airy region
 where the branch split ceases to be local.  Right: the direct multichannel
 calculation verifies the weak-coupling limits in
 Eq.~\eqref{eq:weak_scaling}.}
 \label{fig:new_diagnostics}
\end{figure*}

A direct QR-stabilized integration at $M=12$, $n=2$, and $\omega=24$ gives
\begin{equation}
 C_P=0.054,\qquad C_Q=0.021 .
 \label{eq:weak_coefficients}
\end{equation}
For $0.0017\leq\mu^2\leq0.053$, the normalized quantities
$\Pexc/(C_P\mu^4)$ and $\Delta Q/(C_Q\mu^2)$ approach unity as
$\mu^2\rightarrow0$, as shown in the right panel of
Fig.~\ref{fig:new_diagnostics}.

The matrix Wigner--Smith operator is
\begin{equation}
 \bm Q_M(\omega)=i\bm S_M^\dagger\partial_\omega\bm S_M,
 \quad
 \Qint=\int\dd\omega\,|A|^2
 \bm e_0^\dagger\bm Q_M\bm e_0 .
 \label{eq:matrix_q}
\end{equation}
An incoming expanding section $\alpha_{\rm in}$ and an outgoing contracting
section $\alpha_{\rm out}$ are connected by
\begin{equation}
 \bm U_{\alpha_{\rm out}\leftarrow\alpha_{\rm in}}(\omega)
 =e^{i\omega(\alpha_{\rm in}+\alpha_{\rm out})}\bm S_M(\omega),
 \label{eq:clock_transition}
\end{equation}
so its packet-averaged section duration is
\begin{equation}
 \Qsec=\Qint-\alpha_{\rm in}-\alpha_{\rm out} .
 \label{eq:section_duration}
\end{equation}

\begin{figure*}[t]
 \centering
 \IfFileExists{paper2_figures/fig3_multichannel_structure.pdf}
 {\includegraphics[width=\textwidth]{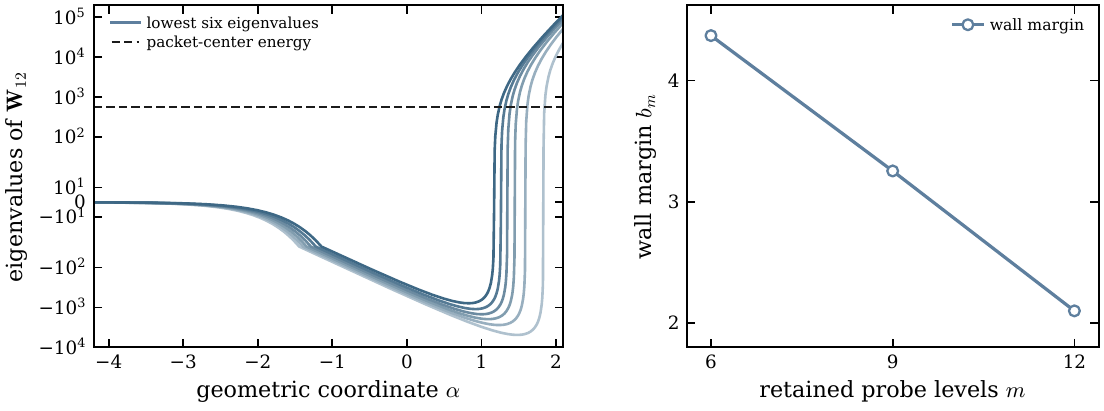}}
 {\missingfigure{Finite-detector potential and nested-level wall margins.}}
 \caption{Finite-detector multichannel structure.  Left: the six lowest
 eigenvalues of the $M=12$ matrix potential and the packet-center energy.
 Right: the wall margin remains positive for the $m=6,9,12$ nested
 projections.}
 \label{fig:multichannel}
\end{figure*}

The benchmark fixes $M=12$, $n=2$, and
\begin{equation}
 \eta=\frac{\mu^2\lambda_{\max}(\bm V_{12})}{6}=0.65,
 \qquad b_{12}=2.10 .
 \label{eq:eta}
\end{equation}
The Gaussian packet has $\omega_0=24$ and
$\sigma_\omega=1.92$.  The calculation gives
$\Pexc=1.01\times10^{-3}$, $\Sent=4.81\times10^{-5}$, and
$\Qint=2.080$.
The outgoing probabilities are
$P_0=0.99$, $P_2=0.001$, and $P_4\sim10^{-9}$, with higher channels
below $10^{-12}$.  The hierarchy is consistent with the selection rule and
with Eq.~\eqref{eq:weak_scaling}. The potential eigenvalues and the positive wall margins of the nested
projections are shown in Fig.~\ref{fig:multichannel}.

The outgoing spectral amplitude also admits two complementary local clock
descriptions.  Scalar-clock conditioning is the covariant transform
\begin{equation}
 \bm\psi_{\rm out}(\tau)
 =\frac{1}{\sqrt{2\pi}}
 \int_0^\infty\dd\omega\,
 \bm c_{\rm out}(\omega)e^{i\omega\tau} .
 \label{eq:scalar_transform}
\end{equation}
Parseval's identity gives
\begin{align}
 \int_{-\infty}^{\infty}\dd\tau\,
 \bm\psi_{\rm out}^\dagger\bm\psi_{\rm out}&=1,\nonumber\\
 \int_{-\infty}^{\infty}\dd\tau\,
 \bm\psi_{\rm out}(\tau)\bm\psi_{\rm out}^\dagger(\tau)
 &=\bm\rho_{\rm p}^{\rm out}.
 \label{eq:parseval_reduction}
\end{align}
Thus scalar conditioning and the spectral partial trace give the same
detector state exactly.

\begin{figure*}[t]
 \centering
 \IfFileExists{paper2_figures/fig4_local_clock_reductions.pdf}
 {\includegraphics[width=\textwidth]{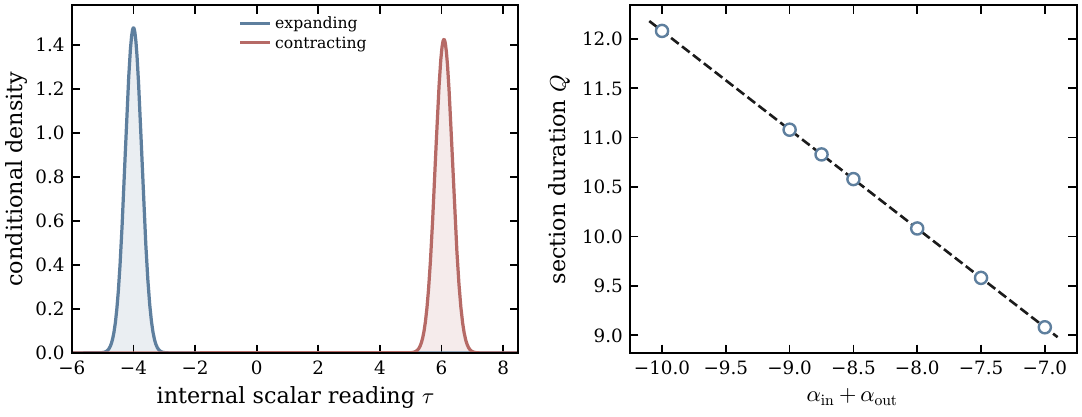}}
 {\missingfigure{Scalar-clock packets and geometric-section covariance.}}
 \caption{Consistency of the two local clock reductions.  Left: incoming
 expanding and outgoing contracting scalar-clock densities.  Right: the
 duration associated with seven oriented section pairs follows
 Eq.~\eqref{eq:clock_covariance}.}
 \label{fig:clock_equivalence}
\end{figure*}

The oriented geometric reduction follows from
Eq.~\eqref{eq:clock_transition}.  Unitarity gives
$\bm U^\dagger\bm U=\I_M$, while the scalar-clock first moment gives
\begin{align}
 \langle\tau\rangle_{\rm out}-\langle\tau\rangle_{\rm in}
 &=\int_0^\infty\dd\omega\,|A(\omega)|^2
 \bm e_0^\dagger
 \left(i\bm U^\dagger\partial_\omega\bm U\right)\bm e_0
 =\Qsec .
 \label{eq:clock_covariance}
\end{align}
The two formulas are concrete local instances of the known equivalence
between Dirac, Page--Wootters, and reduced descriptions: they establish
agreement within the positive-frequency sector selected here. Figure~\ref{fig:clock_equivalence} displays this agreement through the
scalar-clock densities and the section-pair duration.

The numerical transform gives a normalization error
$1.6\times10^{-15}$; the scalar-clock and spectral detector states have trace
distance $6.8\times10^{-16}$, and the two sides of
Eq.~\eqref{eq:clock_covariance} differ by $1.2\times10^{-6}$.  Direct
substitution of the reconstructed packet into
Eq.~\eqref{eq:coupled_wdw} gives normalized stationary residuals
$5.67\times10^{-4}$, $3.61\times10^{-5}$, and $2.27\times10^{-6}$ under two
successive spatial refinements, an empirical order of $3.982$.  Joint
spatial and scalar-clock refinement reaches $2.21\times10^{-6}$ with orders
$3.98$ and $3.97$, respectively.  Appendix~\ref{app:numerics} shows the
definitions and independent stability calculations.

\section{Conclusion}
\label{sec:conclusion}

The principal result is an exact relational duration, where we discard the notion of time as a fundamental physical entity.  In the one-channel model the derivative of the reflection
phase equals the change in the first scalar-clock moment, its variance obeys
Eq.~\eqref{eq:duration_variance}, and its large-frequency limit is the
classical two-crossing duration with the correction
Eq.~\eqref{eq:duration_correction}.  The operational error
Eq.~\eqref{eq:branch_error} then quantifies when an unoriented geometric
reading no longer resolves the two local records.  In the finite-detector
extension the same construction becomes the matrix operator
Eq.~\eqref{eq:matrix_q}; transition probabilities and duration shifts have
different, directly verified weak-coupling orders.

These statements are independent of an external evolution parameter.  The
stationary constrained state contains the incoming and outgoing clock
records, their correlations with the detector, and the scattering phase
needed to reconstruct duration.  Within this model, background time is
therefore dispensable for all reported observables.  Time remains useful as
a relational parameter, but it is supplied by correlations among the
degrees of freedom already present in the constraint.

The reference section in $Q_0(\omega;\alpha_0)$ is physical: it specifies
the two scale-factor crossings whose scalar-clock separation is measured.
The asymptotic channel basis is fixed by the oscillator states and the
covariant scalar POVM.  A common spectral rephasing changes neither the
reflection matrix nor the measured difference of clock moments; allowing
independent energy-dependent rephasings of incoming and outgoing channels
would instead redefine the clock calibration.

Homogeneous minisuperspace is a controlled domain and it does not
contain the local constraint algebra of field theory. The detector is a
finite system rather than a regulator for an unrestricted matter field.
The wall framework is exact for every fixed $M$ satisfying
Eq.~\eqref{eq:wall_theorem}; no $M\to\infty$ claim is needed.  Factor
ordering is another genuine ambiguity.  For the restricted family
\begin{equation}
 [\partial_\alpha^2+q\partial_\alpha-\partial_\tau^2
 -36e^{4\alpha}]\Psi_q=0,
 \label{eq:ordering_family}
\end{equation}
the rescaling $\Psi_q=e^{-q\alpha/2}\Phi_q$ produces the constant shift
$-q^2/4$ and preserves the scattering construction above threshold.
General orderings and lapse rescalings require a separate comparison
\cite{Sahota2024,Kaimakkamis2025,Mondal2025}.

The calculation consequently establishes a precise version of temporal
emergence: crossing order, duration, and detector transitions are recovered
from correlations and scattering data of one stationary Dirac state.  The
result is complete for the stated finite model and supplies an analytically
testable building block for less symmetric constrained systems.

\appendix

\section{Clock identities}
\label{app:proofs}

The decaying Weyl solution of the one-channel Sturm--Liouville problem is
unique at the confining endpoint.  Its generalized spectral representation
is therefore a direct integral with one-dimensional fibers,
\begin{equation}
 \Hphys=\int_{\mathbb R_+}^{\oplus}\mathbb C\,\dd\omega .
\end{equation}
Every bounded operator commuting with $\sqrt{\Theta_0}$ is decomposable; a
commuting projector acts on almost every fiber as either zero or the
identity.  Two such projectors may divide the frequency axis, but they cannot
both select nonzero expanding and contracting states at the same frequency.

For the clock moments, let $c_{\rm out}=S_{\alpha_0}c_{\rm in}$.  The
covariant Fourier transform maps multiplication by $\tau$ to
$T=i\partial_\omega$ on amplitudes that vanish at their support endpoints.
Consequently,
\begin{equation}
 S_{\alpha_0}^\dagger T S_{\alpha_0}=T+Q_0,
 \qquad Q_0=iS_{\alpha_0}^\dagger\partial_\omega S_{\alpha_0}.
 \label{eq:moment_operator_identity}
\end{equation}
Taking first and second moments proves
Eqs.~\eqref{eq:duration_mean}--\eqref{eq:duration_variance}.  For real $A$,
integration by parts gives
$\langle\{\Delta T,\Delta Q_0\}\rangle_A=0$.

The WKB phase in Eq.~\eqref{eq:wkb_branch_phase} obeys
\begin{equation}
 \partial_\omega\vartheta_\pm=\pm\frac{u}{2},
 \qquad
 \partial_\omega^2\vartheta_\pm
 =\pm\frac{1}{2\sqrt{\omega^2-h^2}}.
\end{equation}
Fourier transforming a Gaussian spectral amplitude through quadratic order
therefore gives Eq.~\eqref{eq:branch_clock_width}.  Two equal-variance
Gaussians separated by $u$ have total-variation optimal error
$\tfrac12\erfc[u/(2\sqrt2\sigma_\tau)]$, which gives
Eq.~\eqref{eq:branch_error}.

\section{Finite-detector wall}
\label{app:wall}

The Jacobi matrix $\bm V_M$ is the finite even-sector representation of the
squared oscillator coordinate.  If $x_{M,M}^{(-1/2)}$ denotes the largest
zero of $L_M^{(-1/2)}$, its largest eigenvalue is exactly
\begin{equation}
 \lambda_{\max}(\bm V_M)=\frac{x_{M,M}^{(-1/2)}}{n}.
 \label{eq:laguerre_eigenvalue}
\end{equation}
For $M=12$ and $n=2$ this gives
$\lambda_{\max}=18.09$, so the benchmark in Eq.~\eqref{eq:eta} has
$\mu^2=0.22$ and $b_{12}=2.10$.

Let $x=e^{2\alpha}$ and
$d_M=2M-\tfrac32=\lambda_{\max}(\bm D_M)$.  If $b_M>0$, then
\begin{align}
 \bm W_M(\alpha)
 &\geq(6b_Mx^2-12nd_Mx)\I_M\nonumber\\
 &\geq-\frac{6n^2d_M^2}{b_M}\I_M .
 \label{eq:wall_lower_bound}
\end{align}
Thus the operator is semibounded.  Conversely, if $\bm B_M$ has a negative
direction, the $x^2$ term tends to minus infinity in that direction.  If
$\bm B_M$ has a null direction, the strictly positive matrix $\bm D_M$
makes the remaining $-x$ term tend to minus infinity.  Confinement in every
internal direction is therefore equivalent to $b_M>0$.

The potential is smooth, Hermitian, and limit point at both endpoints.  At
$\alpha\to-\infty$ it tends to zero, whereas at
$\alpha\to+\infty$ its lowest eigenvalue tends to $+\infty$.  Standard
matrix Schr\"odinger results then give the unique self-adjoint realization
and essential spectrum $[0,\infty)$ of multiplicity $M$
\cite{Yafaev1992,Teschl2014}.

\section{Numerical implementation}
\label{app:numerics}

For each $\omega$, a decaying fundamental matrix is initialized at a wall
point where $\bm W_M-\omega^2\I_M$ is positive.  The pair
$(\bm\chi,\bm\chi')$ is propagated toward the open endpoint with adaptive
eighth-order integration.  QR reorthogonalization after short intervals
prevents the independent decaying solutions from becoming numerically
collinear.  From the logarithmic derivative
$\bm Y=\bm\chi'\bm\chi^{-1}$ at $\alpha_L$, the raw reflection matrix is
\begin{equation}
 \bm S_{\rm raw}=-e^{-2i\omega\alpha_L}
 (\bm Y-i\omega\I_M)^{-1}(\bm Y+i\omega\I_M).
 \label{eq:numerical_s}
\end{equation}
Its unitary polar factor is used for observables and centered frequency
differences give $\partial_\omega\bm S_M$.  Moving both numerical endpoints
and tightening the integration tolerance changes the quoted observables
below their displayed precision; the largest raw unitarity defect in the
benchmark grid is $2.4\times10^{-8}$ and the polar correction is of order
$10^{-9}$ in amplitude.

The configuration-space packet is reconstructed as
\begin{equation}
 \bm\Psi(\alpha,\tau)=\int_0^\infty\dd\omega\,
 A(\omega)e^{i\omega\tau}\bm\chi_\omega(\alpha).
\end{equation}
For its Wheeler--DeWitt residual $\bm{\mathcal R}$, the reported normalized
error is
\begin{equation}
 \epsilon^2=\frac{\|\bm{\mathcal R}\|_2^2}
 {\|\partial_\alpha^2\bm\Psi\|_2^2+
  \|\partial_\tau^2\bm\Psi\|_2^2+
  \|\bm W_M\bm\Psi\|_2^2}.
 \label{eq:normalized_residual}
\end{equation}
Extending the spectral support from $\pm6\sigma_\omega$ to
$\pm8\sigma_\omega$ changes $\Pexc$, $\Sent$, and $\Qint$ by
$5.5\times10^{-14}$, $3.6\times10^{-12}$, and $2.7\times10^{-10}$,
respectively.  Nested $M=6,9,12$ calculations agree in the leading
excitation probability to $6.4\times10^{-12}$ on the common retained
channels.  These calculations are independent of the weak-coupling sequence in
Fig.~\ref{fig:new_diagnostics}.

\end{document}